# Temporal evolution of auto-oscillations in a YIG/Pt microdisc driven by pulsed spin Hall effect-induced spin-transfer torque

Viktor Lauer[1], Michael Schneider[1], Thomas Meyer[1], Thomas Brächer[1,*], Philipp Pirro[1], Björn Heinz[1], Frank Heussner[1], Bert Lägel[1], Mehmet C. Onbasli[2], Caroline A. Ross[2,**], Burkard Hillebrands[1,**], and Andrii V. Chumak[1,***]

[1]Fachbereich Physik and Landesforschungszentrum OPTIMAS, Technische Universität Kaiserslautern, 67663 Kaiserslautern, Germany
[2]Department of Materials Science and Engineering, Massachusetts Institute of Technology, Cambridge, Massachusetts 02139, USA
* Present address: Univ. Grenoble Alpes, CNRS, CEA, INAC-SPINTEC, 17, rue des Martyrs 38054, Grenoble, France
** Fellow, IEEE
*** Senior Member, IEEE

*Abstract*— The temporal evolution of pulsed Spin Hall Effect – Spin Transfer Torque (SHE-STT) driven auto-oscillations in a Yttrium Iron Garnet (YIG) | platinum (Pt) microdisc is studied experimentally using time-resolved Brillouin Light Scattering (BLS) spectroscopy. It is demonstrated that the frequency of the auto-oscillations is different in the center and at the edge of the investigated disc that is related to the simultaneous STT excitation of a bullet and a non-localized spin-wave mode. Furthermore, the magnetization precession intensity is found to saturate on a time scale of 20 ns or longer, depending on the current density. For this reason, our findings suggest that a proper ratio between the current and the pulse duration is of crucial importance for future STT-based devices.

## I. INTRODUCTION

The Spin-Transfer Torque (STT) effect [Slonczewski 1996, Berger 1996], which is induced by the Spin Hall Effect (SHE) [Dyakonov 1971, Hirsch 1999] in a heavy metal layer adjacent to a magnetic layer, attracts attention since it can be used for the compensation of magnetization precession damping [Ando 2008, Demidov 2011, Hamadeh 2014, Lauer 2016A] as well as for the excitation of magnetization auto-oscillations [Kajiwara 2010, Demidov 2012, Collet 2016, Demidov 2016] driven by a direct current (see also the review by Chumak et al. [2015]). First successful experiments on the excitation of auto-oscillations were performed on patterned all-metallic bilayers of NiFe/Pt [Demidov 2012] and, later, also on bilayers of the ferrimagnetic insulator Yttrium Iron Garnet (YIG) and Pt [Collet 2016, Demidov 2016]. YIG is known for its very low Gilbert damping and, thus, is important for fundamental research in magnonics and for potential future applications [Serga 2010, Chumak 2015]. Up to now, experiments have been typically performed by applying continuous direct currents to the Pt layer. To the best of our knowledge, only in the experiments of Demidov et al. [2011] the temporal behavior of SHE-STT-driven spin dynamics have been investigated so far. However, the time resolution of 20 ns, which was achieved in these experiments, is larger than the magnon life-time in metallic structures. Thus, it could not provide insight into a very important regime of STT-driven dynamics. In particular, the question how fast the dynamical equilibrium of the auto-oscillations is reached is still open. This is an important aspect since pulsed excitations of the magnetization precession using a pulse duration of a few nanoseconds or shorter are more realistic to the working regime of future spintronic devices. Therefore, time-resolved investigations of the onset of auto-oscillations with a time resolution on the ns-scale are demanded.

Here, we address the experimental investigation of pulsed SHE-STT-driven auto-oscillations (see the schematic of the effects in Fig. 1(a)) in a YIG/Pt microdisc by using time-resolved Brillouin Light Scattering (BLS) spectroscopy [Sebastian 2015] measurements with a time resolution of down to 1 ns. The focus lies on the temporal evolution of the spin dynamics in the microstructure as soon as the anti-damping STT overcompensates the intrinsic Gilbert damping in the system. It is found that the magnetization precession amplitude saturates on a time scale of a few tens of ns depending on the particular current density. Furthermore, both, the maximum intensity and the saturation time, saturate with increasing driving current.

Corresponding author: Andrii V. Chumak (chumak@physik.uni-kl.de).

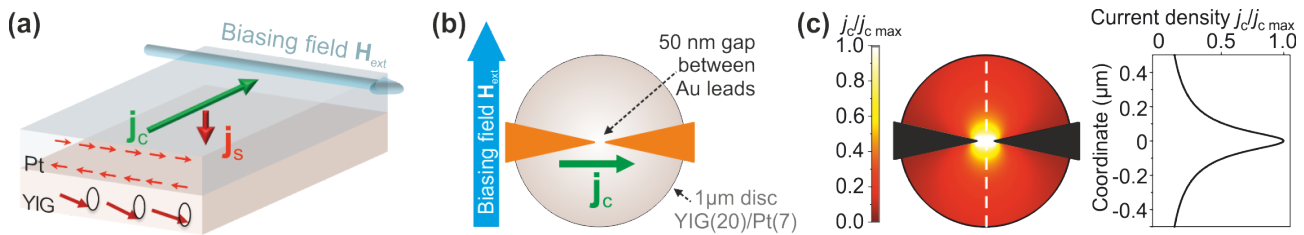

Fig. 1. (a) Configuration of the biasing field, the applied charge and the SHE-STT-generated spin current densities ($\mathbf{j_c}$ and $\mathbf{j_s}$) for the exertion of anti-damping STT on the magnetization in the YIG layer. (b) Sketch of the investigated YIG/Pt disc with tapered Au leads on top. The charge current is applied perpendicular to the biasing field. (c) Current density in the Pt disc calculated using the COMSOL Multiphysics® software. Only the in-plane component of the current that is perpendicular to the biasing magnetic field is taken into account. Right panel shows the cross-section of the current density distribution along the dashed line shown on top of the color map.

## II. STRUCTURE UNDER INVESTIGATION AND METHODOLOGY

The probed YIG/Pt microdisc has a diameter of 1 µm, and the layer thicknesses of YIG and Pt are $d_{YIG}$ = 20 nm and $d_{Pt}$ = 7 nm, respectively – see Fig. 1(b). In order to fabricate the microstructures, a YIG film was grown by pulsed laser deposition on a gadolinium gallium garnet substrate [Onbasli 2014]. Subsequently, after a conventional cleaning in an ultrasonic bath and a treatment of the YIG surface by an oxygen plasma [Jungfleisch 2013], a Pt film was deposited on top by means of sputter deposition. Microwave-based ferromagnetic resonance measurements yielded a Gilbert damping parameter of $\alpha_{YIG}$ = 1.2·10$^{-3}$ for the YIG film before the Pt deposition, and an increased damping value of $\alpha_{YIG/Pt}$ = 5.7·10$^{-3}$ for the YIG/Pt bilayer. The subsequent structuring of the microdisc was achieved by electron-beam lithography and argon-ion milling [Pirro 2014]. Eventually, tapered Au leads on top of the microdisc with a spacing of 50 nm (see Fig. 1(b)) were patterned using electron-beam lithography and physical vapor deposition to allow for the application of high current densities in the center of the disc. The total resistance of the structure under investigation is around 15 Ohm. Since the current density in the Pt disk is not uniform, a numerical simulation was performed using the COMSOL Multiphysics® software. A Pt conductivity value of 2.5·10$^6$ S/m was used in these calculations. The so obtained current density distribution is shown in Fig. 1(c). The in-plane density component that is perpendicular to the biasing field and, thus, contributes to the STT, is plotted in the right panel of Fig. 1(c) as a function of the coordinate along the disc as indicated by the white dashed line on top of the color map. One can see that the density shows a pronounced maximum in the disc between the nano-contacts. Unless otherwise stated, the current density values $j_c$ shown below represent the calculated density maximum in the disc center by taking into account the particular applied voltage, the resistance of the structure, and the density distribution in Fig. 1(c).

Time-resolved BLS measurements were performed by using a probing laser with a wavelength of 491 nm and a power of 2 mW, focused down to a laser-spot diameter of approximately 400 nm on the structure. In the experiment, 75 ns long dc pulses having 5 ns rise and fall times are applied to the Au leads with a repetition period of 500 ns that result in a corresponding charge current flowing in the Pt layer of the microdisc. An external biasing field of $\mu_0 H_{ext}$ = 110 mT magnetizes the microdisc perpendicular to the current flow direction. An exemplary configuration of the biasing field $H_{ext}$ relative to the charge current density $j_c$, and the SHE-STT-generated spin current density $j_s$ are depicted in Fig. 1(a) for the case of a resulting anti-damping STT on the magnetization in the YIG layer [Schreier 2015]. It should be mentioned that, in the present experiment, a threshold-like onset of auto-oscillations for a given field polarity is observed above a critical current density of $j_{c,crit}$ = 1.09·10$^{12}$ A·m$^{-2}$ only for the current direction which is expected to generate an anti-damping STT according to the theory of the SHE. Such a behavior is consistent with other experimental findings as, e.g., shown by Demidov et al. [2012, 2016], Lauer et al. [2016A], and Collet et al. [2016]. Moreover, these observations prove that, unlike in [Safranski 2016, Lauer 2016B], the auto-oscillations cannot originate from the spin Seebeck effect (SSE) due to a thermal gradient, since the SSE is known to excite magnetization precession for both current orientations. In our case, the highly heat-conducting Au leads on top of the Pt layer act as heat sinks and prevent the formation of sufficiently large thermal gradients required for triggering of auto-oscillations due to the SSE. Moreover, the whole structure was covered by a 220 nm thick layer of SiO$_2$ that acts as an additional heat sink and reduces the influence of the sample heating on the studied phenomena.

## III. EXPERIMENTAL RESULTS

Figure 2(a) shows the temporal evolution of the frequency-integrated intensity of the excited magnetization precession represented by the BLS intensity detected in the center of the microdisc for different applied current densities above the critical density value. A dynamic state is apparently excited during the pulse duration of 75 ns illustrated by the shaded areas in the graphs. We find that for current densities higher than 1.58·10$^{12}$ A·m$^{-2}$, the BLS intensity saturates within the pulse duration. Nonlinear magnon scattering processes are assumed to limit a further increase (the interplay between different spin-wave modes excited by the SHE-STT will be reported elsewhere). It is noteworthy that the saturation level is also a function of the applied current density (see Fig. 2(b)). Furthermore, the saturation time (as indicated by the dashed lines in Fig. 2(a)) is plotted in Fig. 2(c). This saturation time drops with the applied current and saturates at a value of approximately 23 ns at high currents.

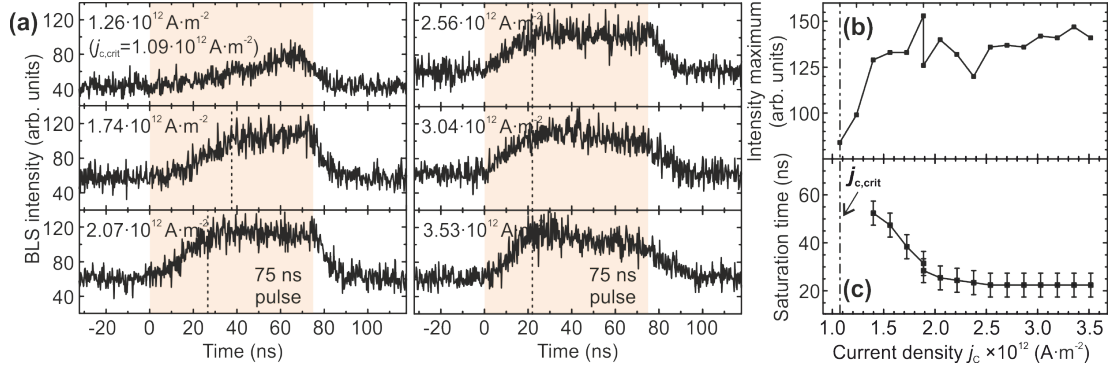

Fig. 2. (a) Frequency-integrated BLS intensity detected in the center of the microdisc as a function of time for different applied current densities above the threshold. The pulse duration of 75ns is depicted by the shaded areas. (b) The maximum BLS intensity reached during the pulse as a function of the applied current density. (c) The saturation time of the BLS intensity (indicated by dashed lines in (a)) within the pulse duration before reaching saturation as a function of the applied current density.

Our findings suggest that the signal output of pulsed auto-oscillations may be optimized with respect to energy consumption by choosing a proper ratio between operating current and pulse duration (compare Figs. 2(b) and (c)). In particular, the importance of the appropriate current value is emphasized by the temporal behavior observed at rather low and rather high currents. For applied voltages that correspond to the current densities below $1.6 \cdot 10^{12}$ A·m$^{-2}$, which are still above the threshold shown in the upper left graph of Fig. 2(a), the pulse duration is too short to reach saturation. On the other hand, for high voltages, e.g., that correspond to the current density $3.5 \cdot 10^{12}$ A·m$^{-2}$ the BLS signal slowly drops within the pulse duration after reaching its maximum after 23 ns, as shown in the lower right graph of Fig. 2(a). This intensity decrease over time within the pulse duration is assumed to result from a decrease in the spin-mixing conductance due to the increase in the total temperature in the microdisc [Uchida 2014], which consequently leads to a reduced injection of the SHE-generated spin current, and to a reduction of the anti-damping STT. Thus, in the integrated structure, the pulse duration should not fall below 23 ns and the optimal operating current density is about $2.1 \cdot 10^{12}$ A·m$^{-2}$ in the center of the disc. These crucial features need to be considered for potential spintronics applications based on pulsed SHE-STT-driven nano-oscillators.

In order to investigate the spatial distribution of SHE-STT-driven spin dynamics, a linescan at an applied current density of $2.1 \cdot 10^{12}$ A·m$^{-2}$ is performed across the disc through the center in the direction perpendicular to the current flow. The BLS intensity integrated during the pulse over all investigated BLS frequencies is plotted in Fig. 3(a) as a function of the position along the disc. It shows a maximum in the disc center, and moderate intensities at the disc edges. Taking into account the current density distribution shown in Fig. 1(c) and the threshold current density of $1.09 \cdot 10^{12}$ A·m$^{-2}$, we conclude that the density of the current at the edges of the disc is not high enough to reach the threshold of auto-oscillations. Nevertheless, the magnetization precession is detected also at the edges of the disc suggesting that a non-localized spin-wave mode is excited in the whole disc by the high current densities in the center. Additionally, the finite size of the laser spot, which is equal to almost half of the disc diameter, should be taken into account and influences the data shown in Fig. 3(a).

In order to better understand the magnetization dynamics in the disc, we have performed frequency-dependent measurements of the signal detected by BLS spectroscopy in the center of the disc and at its edge. Figure 3(b) shows the BLS intensity time-integrated over the 75 ns long pulse as a function of the BLS frequency at different positions on the disc. Please note that the spectra shown in Fig. 3(b) are the result of the subtraction of the spectra with and without applied dc pulses and, thus, all points having intensity larger than zero are associated with STT-based generation of magnetization precession. One can see, that the linewidths of the generated frequency peaks are much larger than the frequency resolution of our BLS setup which is around 50 MHz. Furthermore, the BLS frequency of maximum intensity is lower in the disc center than at the disc edge. We associate this with a simultaneous SHE-STT excitation of at least two modes in the disc that can be simultaneously maintained during the pulse. In the center, a so-called bullet mode is excited that has a solitonic nature and spreads over an area of a few tens of nanometers [Demidov 2012]. This mode is known to have frequencies smaller than the ferromagnetic resonance frequency and is strongly localized in the disc center. Simultaneously, a non-localized mode of higher frequency distributed over the whole disc is excited [Collet 2016, Jungfleisch 2016]. The intensity of this mode is comparable in the disc center and at the edge – see vertical scales in Fig. 3(b). However, the intensity of the bullet mode in the disc center is larger than the intensity of the non-localized mode.

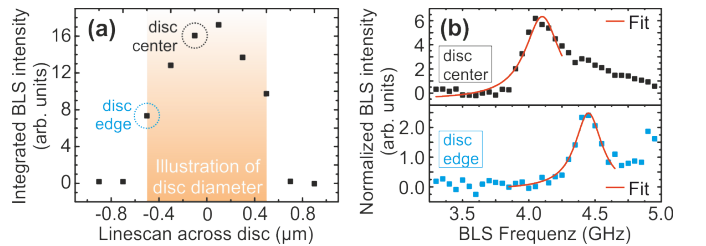

Fig. 3: (a) Linescan perpendicular to the current flow direction showing the time- and frequency-integrated BLS intensity. (b) BLS intensity integrated over the pulse duration as a function of the BLS frequency, normalized to the BLS intensity when the pulse is switched off. The solid lines are Lorentzian fits.

The fact that we see a multi-mode generation, while most of the previously reported studies demonstrate a single-mode generation, leads us to the conclusion that only the application of SHE-STT for a relatively short time period allows for the simultaneous excitation of multiple modes. An application of electric current to the Pt layer for a longer time will probably result in the appearance of nonlinear mode competition mechanisms and in the subsequent survival of only one mode.

## IV. CONCLUSION

In summary, we performed time-resolved BLS measurements to investigate the onset of pulsed SHE-STT-driven magnetization auto-oscillations in a YIG/Pt microdisc. The BLS intensity is found to saturate on a time scale of 25 ns or longer, depending on the particular current density which originates from the applied voltage. Furthermore, both the maximum intensity and the saturation time saturate with increasing operating voltage, which we associate with nonlinear magnon-magnon interaction in the system. For this reason, our findings suggest that a proper ratio between the voltage and the pulse duration is of crucial importance for the power consumption of potential devices based on pulsed auto-oscillations. It was demonstrated that the peak frequency of the SHE-STT-excited magnetization is different in the disc center and at its edges. This is associated with the simultaneous STT excitation of a bullet mode and a non-localized spin-wave mode in the system. This might be attributed to the fact that the system does not reach a quasi-equilibrium state during the first few tens of nanoseconds of an applied current pulse.

## ACKNOWLEDGMENT

This research has been supported by the EU-FET Grant InSpin 612759, and by the ERC Starting Grant 678309 MagnonCircuits.